\documentclass[11pt,preprint]{aastex}

\received{}
\revised{}
\accepted{}

\shorttitle{Galaxy Shape \& Environment}
\shortauthors{Kuehn \& Ryden}

\begin{document}
\title{Dependence of Galaxy Shape on Environment
in the Sloan Digital Sky Survey}
\author{Frederick Kuehn}
\affil{Department of Physics, The Ohio State University,
Columbus, OH 43210}
\email{kuehn@mps.ohio-state.edu}
\and
\author{Barbara S. Ryden}
\affil{Department of Astronomy, The Ohio State University,
Columbus, OH 43210}
\email{ryden@astronomy.ohio-state.edu}

\begin{abstract}

Using a sample of galaxies from the Sloan Digital Sky Survey Data (SDSS) Data Release 4,
we study the trends relating surface brightness profile type and
apparent axis ratio to the local galaxy environment. We use the
SDSS parameter `fracDeV' to quantify the profile type (fracDeV = 1 for a
pure de Vaucouleurs profile; fracDeV = 0 for a pure exponential profile). We find that
galaxies with $M_r \gtrsim -18$ are mostly described by exponential profiles in
all environments. Galaxies with $-21 \lesssim M_r \lesssim -18$ mainly have exponential
profiles in low density environments and de Vaucouleurs profiles in high density
environments. The most luminous galaxies, with $M_r \lesssim -21$, are mostly described
by de Vaucouleurs profiles in all environments. For galaxies with $M_r\lesssim-19$, the fraction
of de Vaucouleurs galaxies is a monotonically increasing function of local density, while the fraction
of exponential galaxies is monotonically decreasing. For a fixed surface brightness
profile type, apparent axis ratio is frequently correlated with environment. As the
local density of galaxies increases, we find that for $M_r \in [-18, -20]$, galaxies of all profile types 
become slightly rounder, on average; for $M_r \in [-20, -22]$, galaxies with mostly exponential profiles
tend to become flatter, while galaxies with de Vaucouleurs profiles tend to become rounder;
for $M_r \in [-22, -\infty]$, galaxies with mostly exponential profiles become flatter,
while the de Vaucouleurs galaxies become rounder in their inner regions, yet exhibit no change
in their outer regions. We comment on how the observed trends relate to the merger history of galaxies.

\end{abstract}

\keywords{galaxies: elliptical and lenticular, cD ---
galaxies: fundamental parameters ---
galaxies: photometry ---
galaxies: spiral ---
galaxies: statistics
}


\section{INTRODUCTION}
\label{sec-intro}

The modern paradigm of hierarchical galaxy formation is rooted in the notion that large stellar systems are formed through the mergers of smaller ones. Understanding the creation and evolution of a particular galaxy requires a knowledge of the merger history of dark halos and the accompanying gas dynamics. Different environments necessarily lead to differing merger histories, resulting in differences in the observable properties of each galaxy. In particular, denser environments lead to more frequent mergers, which cause significant structural changes to galaxies. Also, gravitational harassment among galaxies in dense regions can affect the morphological properties of galaxies. Within the context of the current paradigm of galaxy formation, it is possible to create computer simulations that begin with some matter distribution and then follow the creation and evolution of structure (e.g. see \citet{be98,ad03,be05, rb05}). Observational astronomy, by contrast, is unable to follow the evolution of a particular galaxy. However, statistical correlations between properties such as galaxy color, luminosity, kinematics, and shape have been found. Attempts to understand how such trends are related to the underlying physics of galaxy evolution are a major focus of extragalactic astronomy.  The purpose of this study is to examine both new and well known trends that relate a galaxy's structure to its environment through the use of observable quantities.

Historically, \citet{hu26} classified galaxies by their visual appearance on photographic plates. Elliptical galaxies have smooth elliptical isophotes; spiral galaxies have spiral arms that wind outward from a central bulge or bar. By measuring the surface brightness $I(R)$ along the major axis of a galaxy's image, it was noticed that bright ($M_B\lesssim -20$) ellipticals have light profiles that are well fit by log$\,I$ $\propto$ $-$R$^{1/4}$ \citep{de59}. On the other hand, the azimuthally averaged surface brightness profiles of spiral galaxies were discovered to have log$\,I$ $\propto$ $-$R \citep{fr70}. Such observations reveal a correlation of surface brightness profile with Hubble type. Furthermore, elliptical galaxies are slowly rotating systems characterized by triaxial shapes; in contrast, spiral galaxies are rapidly rotating flattened structures. It has also been found that the average color of galaxies differs for the different galaxy types; ellipticals tend to be redder, while spirals are bluer. Such correlations with Hubble type exist for a host of galaxy properties.

There are well known trends relating morphological type to the local density of galaxies \citep{hh31,dr80,gc03,ta04,pt05}. In general, bright elliptical galaxies are preferentially found in regions of higher galaxy number density; on the other hand, spirals account for the majority of bright galaxies in the field. It is worth noting, however, that such studies have been focused on galaxies with $M_r\lesssim -20$. New, large galaxy surveys open up the possibility of exploring such trends in different magnitude bands.

Images of galaxies are a projection of the true shape. Most studies of the axis ratio of galaxies have concentrated on understanding what the distribution of apparent axis ratios tells us about the intrinsic axis ratios (e.g. \citet{sa70,bv81,la92,el04,vr05}). Such information is a necessary ingredient for, among other things, understanding the kinematic properties of galaxies. Generally, the distribution of apparent axis ratios for thin, nearly circular disks is an approximately uniform function, giving an average axis ratio of $\sim0.5$. For elliptical galaxies, which tend to be triaxial \citep{la92,vr05}, the distribution of apparent axis ratios is a steeply peaked function with an average value of $\sim0.7$. It is also found that the average axis ratio for ellipticals depends on luminosity, with brighter ellipticals being rounder \citep{vr05}. As previously mentioned, bright ellipticals tend to be found in high density environments. This suggests that the eventful merger histories of these systems lead to systematic change in average axis ratio. One under-studied area, related to galaxy shape, is the dependence of axis ratio on environment. \citet{la92} briefly comment on this dependence for elliptical galaxies. However, a continued investigation of this field will help complete understanding of the kinematical properties of galaxies.

In this paper, we use photometric data from the Sloan Digital Sky Survey Data Release 4 (SDSS DR4) \citep{am05} to study the correlations of two structural properties of galaxies with environment; namely, surface brightness profile and apparent axis ratio. The SDSS provides an unprecedentedly large data set with a uniform data reduction scheme, allowing for a consistent study. SDSS data has already been used to quantify correlations between many galaxy properties \citep{bl03,gc03,hg03,bl04,hg04,ta04}.

In \S\ref{sec-data} of this paper, we describe the SDSS, our morphological classification scheme, and the methods by which we determine the apparent axis ratio of galaxies. In \S\ref{sec-prof}, we present our results on the surface brightness profile type -- environment relationship. We make two extensions to previous studies. First, we use galaxies in the relatively broad absolute magnitude range $M_r \in [-17,\, -23]$. Second, we subdivide the data into narrow magnitude bins to determine the luminosity dependence of the profile type -- environment relation. In \S\ref{sec-axis}, we describe the results of our study of the apparent axis ratio -- environment correlation. Motivation for this research was initiated by a prior analysis of galaxy axis ratios in the SDSS data \citep{vr05}, where it was proposed that axis ratio could be correlated with environment. In \S\ref{sec-disc} we place our results in the context of the current understanding of galaxy formation and evolution.


\section{DATA}
\label{sec-data}

The Sloan Digital Sky Survey \citep{yo00, st02} will, when complete, 
provide a map of nearly one-fourth of the celestial sphere. A
CCD mosaic camera \citep{gu98} images the sky in five
photometric bands ($ugriz$; \citet{fu96, sm02}). The follow-up
spectroscopic survey provides spectra for galaxies to a limiting
Petrosian magnitude $m_r = 17.77$. The Sloan Digital Sky Survey
Data Release 4, released to the astronomical community
in 2004 October, covers a photometric area of $6670 {\rm\,deg}^2$
and a spectroscopic area of $4783 {\rm\,deg}^2$
(\citet{am05}; see also \citet{st02}, \citet{ab03}, \citet{ab04}, and \citet{ab05}).

The SDSS DR4 data processing pipeline provides a morphological
star/galaxy separation, with extended objects being classified
as `galaxies' and unresolved objects being classified as `stars'.
For each galaxy, in each photometric band, two models are
fitted to the two-dimensional galaxy image. The first model
has a de Vaucouleurs surface profile \citep{de48}:
\begin{equation}
I(R) = I_e \exp \left( -7.67 [ (R/R_e)^{1/4} - 1 ] \right) \ ,
\end{equation}
which is truncated beyond $7 R_e$ to go smoothly to zero
at $8 R_e$. The second model has an exponential profile:
\begin{equation}
I(R) = I_e \exp \left( -1.68 [ R/R_e - 1 ] \right) \ ,
\end{equation}
which is truncated beyond $3 R_e$ to go smoothly to zero
at $4 R_e$. For each model, the apparent axis ratio $q_m$
and the phase angle $\varphi_m$ are assumed to be constant
with radius. The parameters $q_m$, $\varphi_m$, $R_e$, and
$I_e$ are varied to give the best $\chi^2$ fit to the
galaxy image, after convolution with a double-Gaussian
fit to the point spread function.

A further fit to each galaxy is made by taking the best
de Vaucouleurs model and the best exponential model,
and finding the linear combination of the two that
gives a new best fit.
The fraction of the total flux contributed by the
de Vaucouleurs component is the parameter fracDeV,
which is constrained to lie in the interval $[0,1]$.
The fracDeV parameter is functionally
equivalent to the \citet{se68} index $n$ in
the interval $1 \leq n \leq 4$ \citep{vr05}.
If a galaxy has a S\'ersic index $n$, then
$n = 1$ corresponds to fracDeV = 0,
$n = 4$ corresponds to fracDeV = 1, and the dependence
of fracDeV upon $n$ is monotonic in the interval
$1 \leq n \leq 4$. Following \citet{vr05}, we
call galaxies with $0 \leq$ fracDeV $\leq 0.1$
($n \lesssim 1.2$) \textit{ex} galaxies. Galaxies
with $0.1 \leq$ fracDeV $\leq 0.5$ ($1.2 \lesssim n
\lesssim 2.0$) are labeled \textit{ex/de} galaxies.
Galaxies with $0.5 \leq$ fracDeV $\leq 0.9$
($2.0 \lesssim n \lesssim 3.3$) are labeled \textit{de/ex}
galaxies. Finally, galaxies with fracDeV $\geq
0.9$ ($n \gtrsim 3.3$) are labeled \textit{de} galaxies.

The SDSS DR4 databases provide many different measures of the
apparent axis ratio $q$ of each galaxy in each of the five
photometric bands. In this paper, we use the $r$ band data,
at an effective wavelength of 6165{\AA}; this is the band
in which the star/galaxy classification is made.
A useful measure of the apparent shape in the outer regions of galaxies
is the axis ratio of the 25 mag arcsec$^{-2}$ isophote. The SDSS DR4
data pipeline finds the best fitting ellipse to the 25 mag arcsec$^{-2}$
isophote in each band; the semimajor axis and semiminor axis of this
isophotal ellipse are $A_{25}$ and $B_{25}$. The isophotal axis ratio
$q_{25} \equiv B_{25}/A_{25}$ then provides a measure of the apparent
galaxy shape at a few times the effective radius. For galaxies in
our sample with fracDeV = 1,
$A_{25} \sim 3.2 R_e$; for galaxies with fracDeV
= 0, $A_{25} \sim 2.4 R_e$.

The second measure of the apparent axis ratio that we
use is $q_{\rm am}$, the axis ratio determined by the use
of adaptive moments of the galaxy's surface brightness. The
method of adaptive moments determines the $n$th order moments of a
galaxy image, using an elliptical weight function whose shape matches
that of the image \citep{bj02,hs03}. The SDSS DR4 adaptive moments use
a weight function $w (x,y)$ that is a Gaussian matched to the size and
ellipticity of the galaxy image $I(x,y)$. The adaptive first order moments,
\begin{equation}
\vec{x}_0 = {\int \vec{x} w(x,y) I(x,y) dx dy \over \int w(x,y) I(x,y) dx dy} \ ,
\end{equation}
tell us the `center of light' of the galaxy's image. With this knowledge,
we can compute the adaptive second order moments:
\begin{equation}
M_{xx} = {\int (x-x_0)^2 w(x,y) I(x,y) dx dy \over \int w(x,y) I(x,y)
dx dy} \ ,
\end{equation}
and so forth. The SDSS DR4 provides for each image the values of the
parameters $\tau = M_{xx} + M_{yy}$, $e_+ = (M_{xx}-M_{yy})/\tau$,
and $e_\times = 2 M_{xy} / \tau$. The adaptive second moments can
be converted into an axis ratio using the relation
\begin{equation}
q_{\rm am} = \left( {1-e \over 1+e} \right)^{1/2} \ ,
\end{equation}
where $e = (e_+^2 + e_\times^2)^{1/2}$.
The adaptive moments axis ratio $q_{\rm am}$ is not corrected for the effects of seeing.
The SDSS DR4 also provides the fourth order adaptive moments of the
galaxy image, and the adaptive moments $\tau_{\rm psf}$, $e_{+,{\rm psf}}$,
and $e_{\times,{\rm psf}}$ of the point spread function at the galaxy's
location. These moments can be used to correct for the smearing and
shearing due to seeing; such corrections are essential for studying
the small shape changes resulting from weak lensing \citep{bj02,hs03}.
However, when we examine the apparent axis ratios of galaxies
in Section~\ref{sec-axis}, we will only look at well-resolved
galaxies ($\tau > 6.25 \tau_{\rm psf}$), for which the
seeing corrections are negligible. We note that, at most, weak lensing shears the shape of a galaxy at the one percent level \citep{sh04}. Such minute changes in the shape do not significantly affect our results.

Our complete sample of galaxies consists of those objects in the SDSS DR4
spectroscopic sample which are flagged as galaxies, which have
$\tau \geq \tau_{\rm psf}$, and which have spectroscopic
redshifts $z > 0.004$ and a redshift confidence parameter
$zconf > 0.35$. The absolute magnitude $M_r$ of each galaxy is computed
from its SDSS cmodel apparent magnitude, assuming a uniform Hubble
flow with $H_0 = 70 {\rm\,km}{\rm\,s}^{-1}{\rm\,Mpc}^{-1}$, $\Omega_{M,0}
= 0.3$, and $\Omega_{\Lambda,0} = 0.7$. The complete sample contains
$N = 305 \, 558$ galaxies with a median redshift of $z \sim 0.1$. The morphology
of each galaxy in the sample is quantified by the parameter fracDeV,
and by the two measures of the apparent axis ratio: $q_{25}$, which
gives the shape in the outer region, and $q_{\rm am}$, which is
more strongly weighted toward the inner region of the galaxy.


\section{PROFILE TYPE}
\label{sec-prof}


\subsection{SURVEY VOLUME}\label{sec-vol}

A galaxy with absolute magnitude $M_r$ will enter the SDSS spectroscopic survey if its luminosity distance is less than $d_L=10^{0.2(m_r-M_r-25)}$ Mpc, where the limiting apparent magnitude of the survey is $m_r=17.77$. In our study of the relationship between surface brightness profile type and environment, we create six volume-limited samples with absolute magnitude cut-offs $M_{r_0}$ = $-17$, $-18$, $-19$, $-20$, $-21$, and $-22$. The maximum luminosity distance in the volume-limited samples ranges from
$d_L = 10^{1.954} {\rm\,Mpc} = 90 {\rm\,Mpc}$ for the $M_{r_0} = -17$ sample (corresponding to $z_{\rm max} \approx 0.02$),
to $d_L = 10^{2.954} {\rm\,Mpc} = 900 {\rm\,Mpc}$, for the $M_{r_0} = -22$ sample (corresponding to $z_{\rm max} \approx 0.2$).
In each sample we choose as our `target galaxies' those galaxies with $M_r \in [M_{r_0},\; M_{r_0}-1]$. The target
galaxies are those galaxies for which we will determine the profile type -- environment dependence. The total number
of target galaxies in each volume-limited sample, and the fraction of target galaxies corresponding to each of our
four morphological types, is presented in Table~\ref{tab:profile}. 
Table \ref{tab:profile} also lists, under the heading `cylinder width', the total number of cylinders (see \S\ref{sec-env}) that do not intersect the border of the SDSS spectrographic survey, which is the same as the number of galaxies that we keep in our analysis (see \S\ref{sec-env}). The galaxies in each sample are subject to the quality cuts described in \S\ref{sec-data}.

\subsection{THE ENVIRONMENT}\label{sec-env}

The environment of each target galaxy is determined by a count of neighboring galaxies. We adhere to the following in determining this count. A cylinder is created around each target galaxy in redshift space, with the long axis of the cylinder lying
along the line of sight.  The cylinder stretches 6 $h^{-1}$ Mpc (with $h=0.7$) in either radial direction from the target galaxy (i.e. 12 $h^{-1}$ Mpc in total). The radius of the cylinder is one of the following: 0.5 $h^{-1}$ Mpc, 2 $h^{-1}$ Mpc, 8 $h^{-1}$ Mpc. Counting the number of galaxies with $M_r \leq M_{r_0}$ in this cylinder defines our density parameter. We note that in dense environments, this method undercounts the true number of galaxies in each cylinder. This is because the spectroscopic fibers on each SDSS plate cannot be placed more closely than $55\arcsec$. This causes $\sim 7\,\%$ of galaxies (of the total) to be lost \citep{bl03}, undercounting galaxies in high density regions. Fibers are assigned to galaxies with no correlations to galaxy properties \citep{blli03}. Hence, there should be no significant effect to our results. For example, if $7\,\%$ of each galaxy type in the highest density bin were moved to the second-highest density bin, the relative fraction would not change. In general, this is the largest source of galaxy loss, as the completeness of the spectroscopic survey is $\sim 91\%$ \citep{bl03}.

\begin{center}
\begin{figure}
\plotone{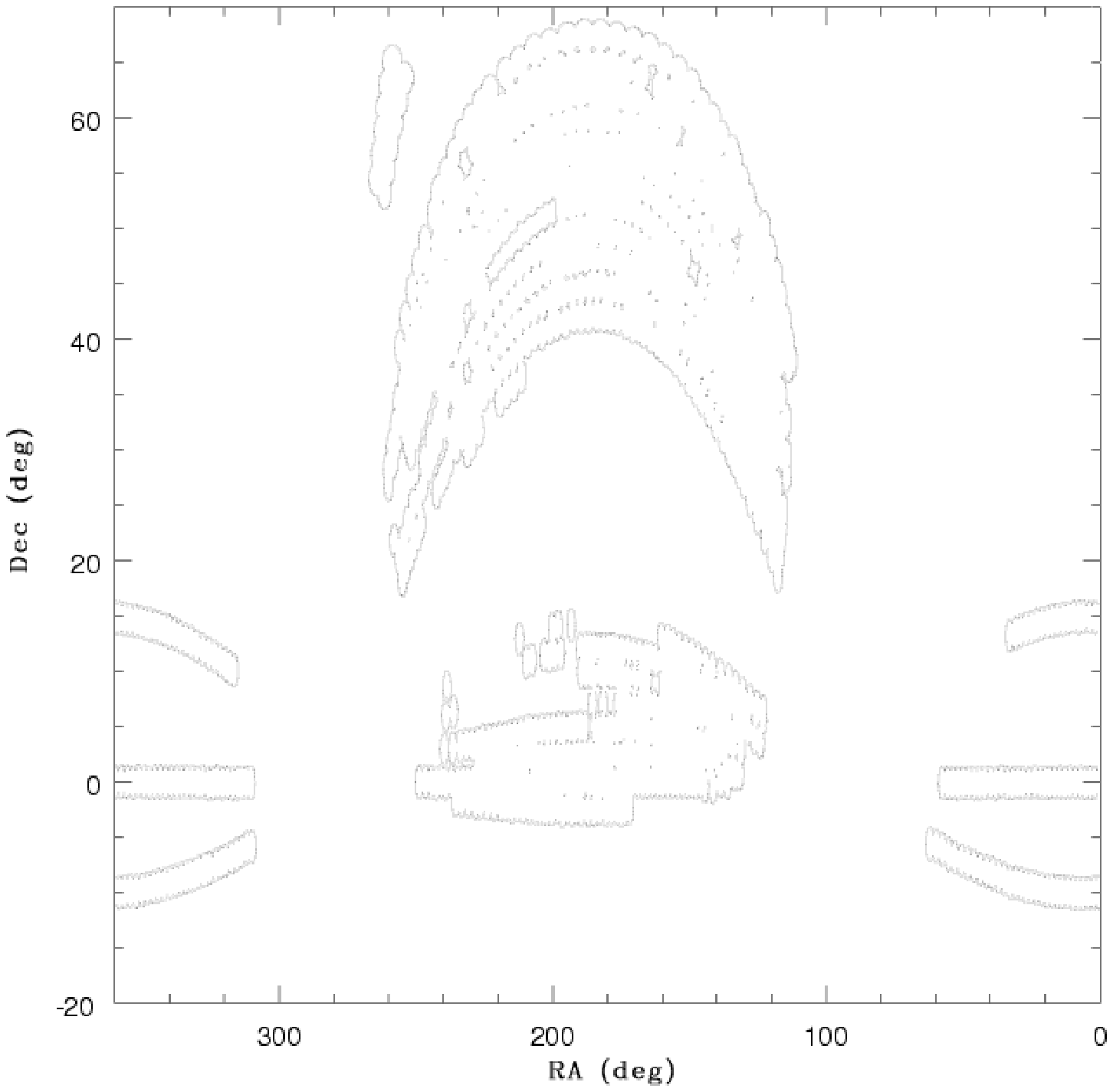}
\caption{Boundary of the SDSS DR4 spectroscopic survey.\
}
\label{fig:1}
\end{figure}
\end{center}

The SDSS takes spectra using circular plates whose diameter subtends an angle of $3^\circ$ on the sky.
The SDSS DR4 spectroscopic survey does not completely match the area covered by the DR4 photometric
survey. Figure~\ref{fig:1} shows the boundary of the spectroscopic survey. If a target galaxy is within
one cylinder width of the border, it is dismissed from our analysis. In addition, if a galaxy is
within one-half cylinder length, $6 h^{-1} {\rm\,Mpc}$, of the maximum luminosity distance, it is
dismissed from our analysis.

In determining the environment for target galaxies in each magnitude bin we count \emph{all} of the galaxies brighter than the magnitude cut-off $M_{r_0}$. In plotting our results, we group together at least 100 galaxies in each bin. We take the weighted mean of the number of galaxies in each cylinder that falls into a particular bin. This defines the local density, or environment for all galaxies in each bin. The highest density bin may have as few as 50 galaxies (the only excpetion to this rule is the case with $M_{r_0}=-17$ and $2\, h^{-1}$ Mpc cylinders where the number in each bin is half that just described). We use the same density bins for all profile types so that we may determine the relative fraction of each. The bins are determined by the binning of the \textit{ex} galaxies subject to the previously given conditions. We changed the possible binning for this analysis, varying it by profile type, and found no substantial change in the results.


\subsection{RESULTS FOR PROFILE TYPE -- ENVIRONMENT RELATIONSHIP}
\label{sec-profresults}

Many studies (e.g. \citet{dr80,gc03,pt05}) have shown that as the density of galaxies in a region increases, the relative fraction of elliptical galaxies to spiral galaxies increases. Instead of using the Hubble classification scheme, we classify galaxies based on their surface brightness profile type, using four classes ranging from the highly concentrated \textit{de} galaxies
to the least concentrated \textit{ex} galaxies.

Our largest survey volume extends to the distance at which a galaxy with $M_r=-22$ has $m_r=17.77$, corresponding to $z\approx 0.2$. \citet{nu05} have shown that in the redshift interval $0 < z < 0.2$, there is little evolution in the fraction of galaxies
with $n > 2$ (corresponding to our classes \textit{de} and \textit{de/ex}). Since all our other magnitude cut-offs correspond to lower maximum redshifts, we do not expect significant redshift evolution effects in our samples.  

\begin{figure}
\plotone{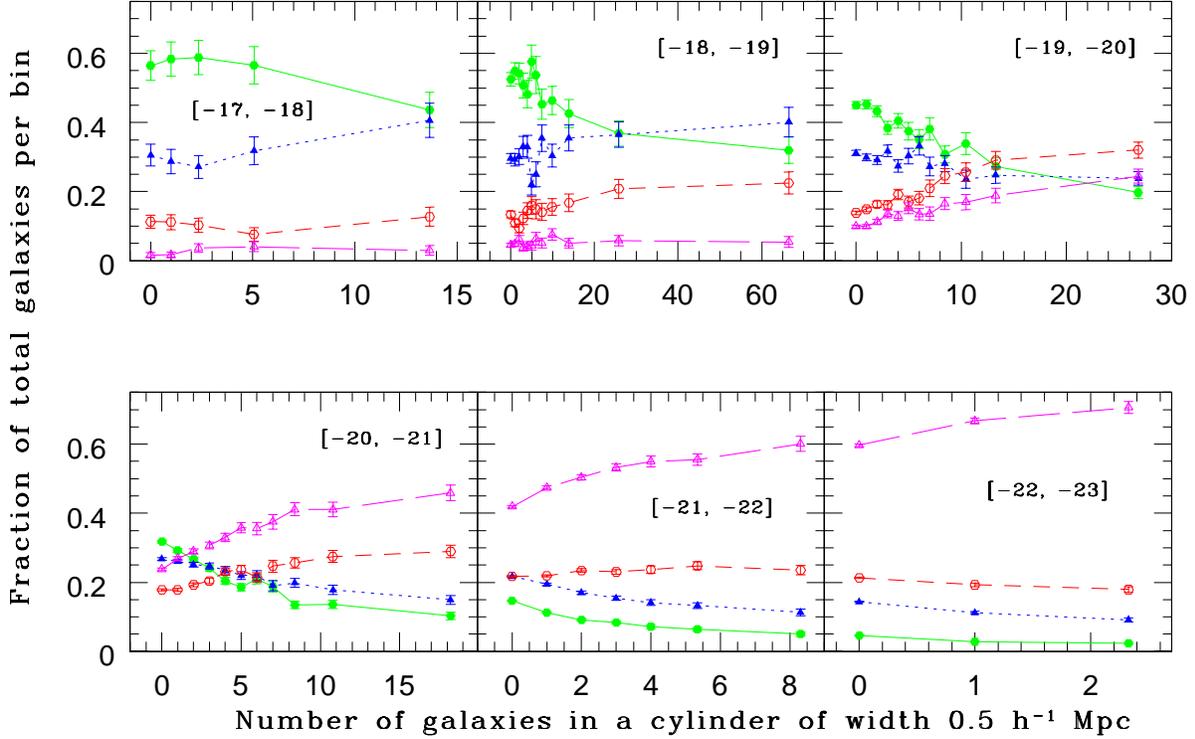}
\caption{Profile type -- environment relation, using cylinders of width 0.5 $h^{-1}$ Mpc. Panels display the fraction of each surface brightness profile type vs. the number of local galaxies for the magnitude ranges [-17, -18], [-18, -19], [-19, -20], [-20, -21], [-21, -22], and [-22, 23]. Solid, green lines with filled circles indicate \textit{ex} type galaxies; dotted, blue lines with filled triangles are for \textit{ex/de} type galaxies; short dashed, red lines with open circles are for \textit{de/ex} type galaxies; long dashed, magenta lines with open triangle are for \textit{de} type galaxies.
}
\label{fig:2}
\end{figure}

\begin{figure}
\plotone{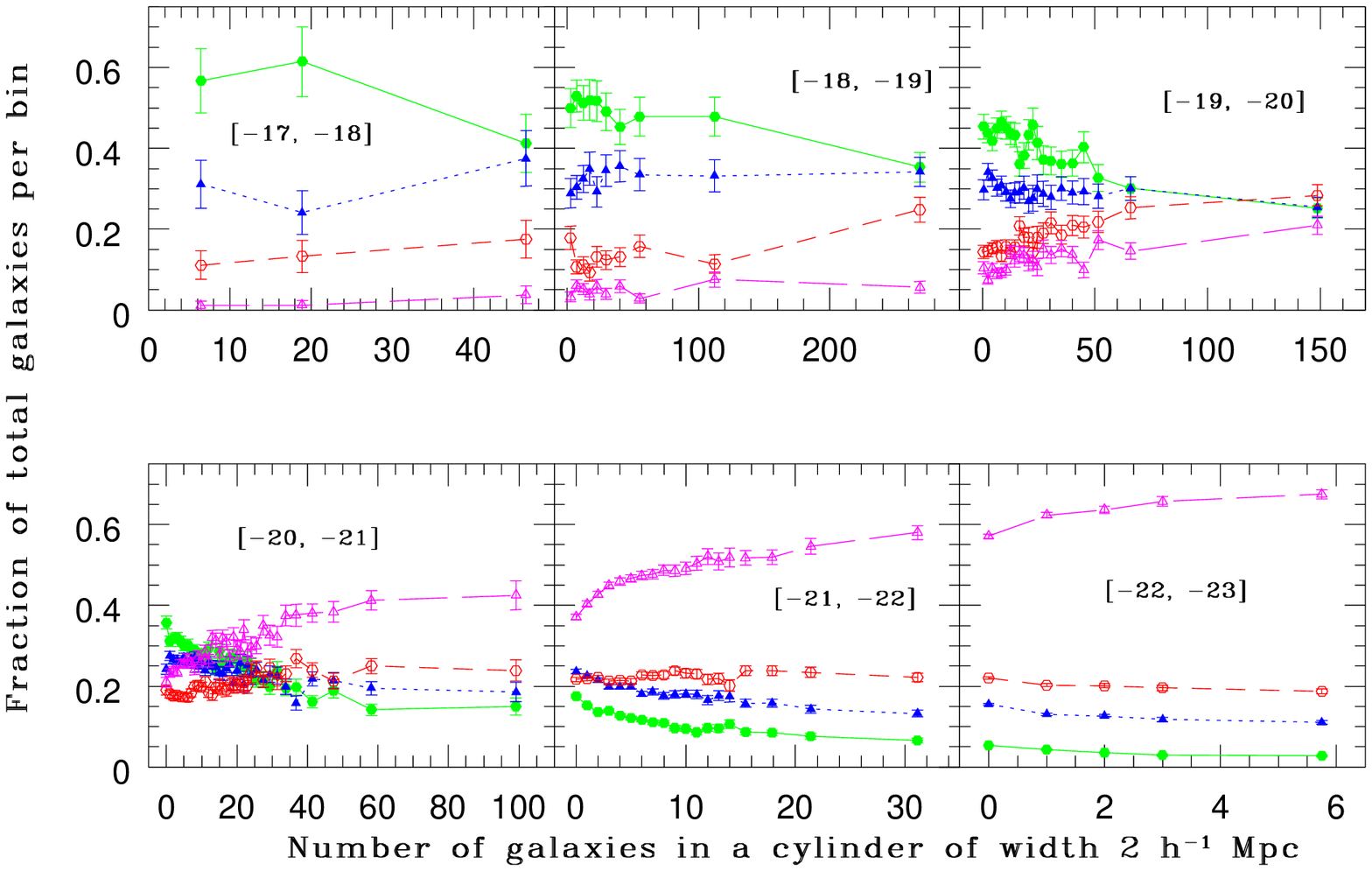}
\caption{Same as Figure~\ref{fig:2}, but with cylinders of width 2 $h^{-1}$ Mpc.
}
\label{fig:3}
\end{figure}

\begin{figure}
\plotone{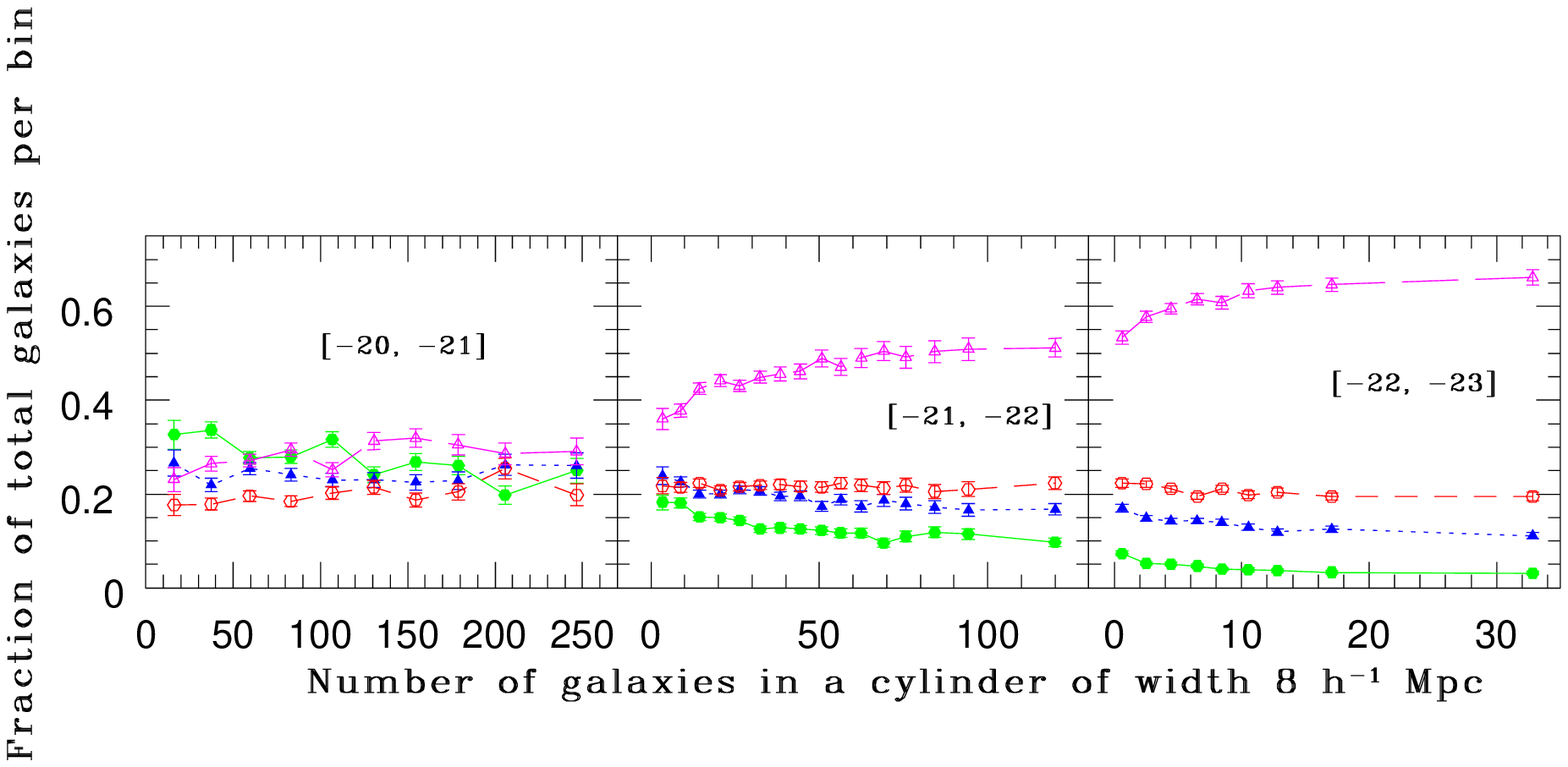}
\caption{Same as Figure~\ref{fig:3}, but with cylinders of width 8 $h^{-1}$ Mpc.
}
\label{fig:4}
\end{figure}

Figures \ref{fig:2} through \ref{fig:4} display our findings on the trends of galaxy profile with environment. The plots are organized by magnitude cut-off and cylinder radius.  The \textit{ex} galaxies are represented by green solid lines and solid circles; the \textit{ex/de} galaxies, by blue dotted lines and solid triangles; the \textit{de/ex} galaxies, by red dashed lines and open circles; the \textit{de} galaxies, by magenta long-dashed lines and open triangles. The plotted errors are the expected errors in the mean, assuming Poisson statistics. Table \ref{tab:profile} lists the number of galaxies in each of our volume-limited surveys. It also lists the fraction of each profile type. Finally, it lists the total number of galaxies that survive all of our quality cuts (see \S\ref{sec-data} and \S\ref{sec-env}).

We summarize the results as follows:

1. Figure \ref{fig:2} displays the results for cylinders of width $0.5\, h^{-1}$ Mpc. There are four primary results. First, galaxies with $M_r\gtrsim -19$ are found to have mostly exponential light profiles in all environments. Second, galaxies with $M_r\lesssim -21$ have mostly de Vaucouleurs light profiles in all environments. Third, for galaxies with $M_r\lesssim-19$, as local density increases, the fraction of de Vaucouleurs galaxies increases monotonically with increasing local density, while the fraction of exponentials decreases. Fourth, there is a crossover of predominant profile type between high and low density environments only for galaxies with $M_r\sim -20$. Plotting results for so many magnitude bins clearly shows how the change in dominant profile type depends not only on environment, but also on absolute magnitude. \citet{dr80} examined the morphology -- environment relation for galaxies with $M_V < -19.67$ (assuming $h=0.7$); for a typical color $B-V \approx 0.9$, this corresponds roughly to 
$M_r \lesssim -20$.

2. Figures \ref{fig:3} and \ref{fig:4} display the effect of increasing the cylinder size. The main result is that trends are washed out on scales larger than about 2 $h^{-1}$ Mpc. There is no plot for the magnitude cut-offs $-17$, $-18$, and $-19$ for cylinders that are $8\, h^{-1}$ Mpc wide; the volume of space probed is small enough that once all of the cylinders that intersect with the border are thrown out, there are too few galaxies left over to make the plot statistically interesting.


\section{APPARENT AXIS RATIO}
\label{sec-axis}

This section of the paper addresses how the average shape of galaxies, as quantified by their apparent axis ratio, depends on their environment, for a fixed luminosity and profile type. Our analysis of this topic is described below. We first highlight any differences in the methodology from the preceding section. Unless otherwise mentioned, our methods here remain the same as those described in \S\ref{sec-prof}.

One idiosyncrasy discovered in \citet{vr05} is that for fixed luminosity the average axis ratio is not always a monotonic function of S\'ersic index. Subdividing all galaxies into only two groups, one with all galaxies with fracDeV $<$ 0.5, or $n\lesssim 2$, and another with the remaining galaxies, would average over this interesting behavior. This provides us with further motivation for maintaining our galaxy classification scheme, where we place galaxies in one of the four groups: \textit{ex}, \textit{ex/de}, \textit{de/ex} and \textit{de} (see \S\ref{sec-data}).


\subsection{SURVEY VOLUME}
\label{sec-vol2}

The survey volume for this analysis is similar to that described in \S\ref{sec-vol}. An important difference is that we place a stricter cut on the point spread function (see \S\ref{sec-data}) and in the process lose many de Vaucouleurs galaxies. In \citet{vr05} it was noted that there is a luminosity -- axis ratio trend. In the range $[-18, -20]$ the trend is approximately flat. In the range $[-20, -22]$ the axis ratio increases with increasing luminosity. For galaxies brighter than $-22$, the axis ratio seems to level off. In principle, this correlation could affect our results. This motivates us to keep luminosity bins as narrow as possible, so as to minimize any effect of this trend within a single bin. On the other hand, statistical claims are only useful for large sample sizes. To keep our subsamples of galaxies large enough, we are motivated to keep luminosity bins as wide as possible. As a compromise, we make luminosity bins two magnitudes wide.

We start by creating three volume-limited surveys with absolute magnitude cut-offs $M_{r_0} =  -18, -20,$ and $-22$. For the volumes corresponding to the first two cut-offs, we determine the environment (see \S\ref{sec-env}) for target galaxies that have absolute magnitude $M_r \in [M_{r_0}, M_{r_0}-2]$, while for the brightest cut-off we find the environment for all galaxies with $M_r \in [-22, -\infty]$. There are only sixteen galaxies that are brighter than $-24$ in absolute magnitude in our volume; therefore, their inclusion does not significantly widen the effective magnitude range of the bin. For example, in the volume corresponding to the $M_{r_0}=-18$ cut-off we find the environment for all galaxies with $ M_r \in [-18,-20]$ where the luminosity distance of a galaxy is no larger than $d_L=$ 143 Mpc, corresponding to the maximum luminosity distance at which a galaxy of absolute magnitude $M_{r_0}=-18$ could be observed. The total number of galaxies in each volume, and the percentage of each profile type is presented in Table \ref{tab:shape}. Table \ref{tab:shape} also lists, under the heading `cylinder width', the total number of cylinders (see \S\ref{sec-env}) that do not intersect the border of the SDSS spectrographic survey, which is the same as the number of target galaxies that we keep in our analysis. The galaxies in each sample are subject to quality cuts which are described in \S\ref{sec-data}. The binning procedure is described in \S\ref{sec-env}.


\subsection{RESULTS FOR THE APPARENT AXIS RATIO -- ENVIRONMENT RELATIONSHIP}
\label{sec-shaperes}

Figures \ref{fig:5} through \ref{fig:7} show the correlations of apparent axis ratio with environment, labeled by profile type. Tables \ref{tab:shape18} through \ref{tab:shape22} contain all related statistical information. We plot a best fit straight line through the binned data that minimizes the chi-square. The value of the reduced chi-square for each fit is printed in the tables. The error bars show the error in the mean assuming Poisson statistics. The choice of colors, line type and point type are the same as previously described in \S\ref{sec-profresults}.

Since the average axis ratio depends on the average luminosity, the normalization of each  best fit line, for the axis ratio -- environment trend, is set by the average luminosity for that profile type. Also, since the luminosity functions for each profile type are different in the different volume-limited samples, the average luminosity of each profile type is slightly different. For example, in the $M_r\in [-20,-22]$ range, the \textit{ex} galaxies have an average magnitude of $M_r\approx -20.5$, while the \textit{de} galaxies have $M_r\approx -21.3$. From the luminosity -- axis ratio trend, it is expected that the \textit{de} galaxies would be rounder. This is reflected in Figure \ref{fig:6}, say, where the trend for the \textit{de} galaxies lies above the trend for the \textit{ex} galaxies. Therefore, while it is useful to compare the difference in the axis ratio as determined by the 25 mag arcsec$^{-2}$ isophote versus the axis ratio as determined by the adaptive moments technique, comparisons of axis ratios between galaxy profile types need to be made with the caveat that they contain different luminosity information.

In the absolute \textit{r} band magnitude range $[-20, -22]$, it has been shown \citep{vr05} that, for all profile types, there is a trend toward roundness as galaxies become brighter. In particular, the average axis ratio may increase by as much as approximately 0.05 when the magnitude decreases by unity \citep{vr05}. However, we find that the luminosity in each local density bin is equal to within one tenth of a magnitude confirming that any trend we see are not due to luminosity. Specifically, for $M_r \in [-20, -22]$, the average magnitude for the \textit{ex} galaxies is $-20.5$, for the \textit{ex/de} galaxies it is $-20.7$, for the \textit{de/ex} galaxies it is $-21.0$ and for the \textit{de} galaxies it is $-21.3$. In other magnitude ranges trends of mean axis ratio with luminosity are weaker and we also find that magnitude does not vary significantly from bin to bin.

Table \ref{tab:shape} lists the number of galaxies in each of our volume-limited samples. Also, it lists the fraction of galaxies of each profile type. Tables \ref{tab:shape18} through \ref{tab:shape22} list all statistical information related to each profile type and cylinder size: the chi-square of the best fit line; whether the axis ratio is increasing (`+') or decreasing (`-') with increasing local density; our statistic labeled `significance' is the ratio of the slope to its standard deviation, characterizing if the slope is significantly different than zero.

We summarize the results.

1. In the magnitude range $[-18, -20]$ the data show one statistically important trend at the $3\,\sigma$ level. The \textit{ex/de} galaxies get rounder as the local density of galaxies increases, and the axis ratio, as measured by the axis ratio determined by the moments, for the \emph{de/ex} galaxies increases with density. Other profile types in this magnitude range tend to become rounder at higher local density, but at a lower level of statistical significance. Plots are only created for cylinders of width 0.5 $h^{-1}$ Mpc and 2 $h^{-1}$ Mpc. When cylinders of width 8 $h^{-1}$ Mpc are considered, too many are near the border leaving a statistically uninteresting sample. 

2. In the magnitude range $[-20, -22]$ we find the strongest trends between axis ratio and environment. Generally, the \textit{de} and \textit{de/ex} galaxies get rounder with increasing local density. This trend is more pronounced in the inner regions of these galaxies. On the other hand, the \textit{ex/de} galaxies tend toward greater flattening as the local density of galaxies increases, with a similar trend, though less statistically significant, observed for the \textit{ex} galaxies. Trends of axis ratio with environment are only significant on scales less than or equal to 2 $h^{-1}$ Mpc.

3. For our sample of the brightest galaxies, in the magnitude range $[-22, -\infty]$, we find that trends are more pronounced in the inner regions than the outer regions; trends in the shape, determined by the moments of the light distribution, are no longer present when we use the axis ratio determined by the 25 mag arcsec$^{-2}$ isophote. This is especially true of the \textit{de} galaxies. Statistically, they show a strong positive trend in their axis ratio as determined by the moments of their light distribution. However, the average axis ratio for these galaxies derived from the 25 mag arcsec$^{-2}$ isophote is essentially constant with environment.

\section{DISCUSSION}
\label{sec-disc}

The novel aspect of our study of the surface brightness profile type -- environment relationship is our use of a large magnitude range subsequently divided into many volume-limited samples. Averaging over a wide magnitude range blurs out some of the observed trends, resulting in an inability to distinguish them. Using many volume-limited samples allows us to determine that, at absolute \textit{r} band magnitudes around $-17$, galaxies are almost all described by purely exponential light profiles at all densities, with no strong trends in the fraction of different profile types with environment. For absolute magnitudes around $-19$ and $-20$ there is a crossover from low densities to high densities; galaxies with exponential profiles are more common at low densities, while at high local densities galaxies are most likely to be well described by a de Vaucouleurs profile. Finally, for bright galaxies with $M_r \sim -22$, most galaxies have de Vaucouleurs profiles. 

These results fit well into the hierarchical structure formation picture. Generally, the result of mergers is to increase entropy. In spiral galaxies, stellar orbits in the disk are nearly circular and all of the disk stars travel in the same direction, corresponding to a low overall entropy. On the other hand, orbits of stars in elliptical galaxies are unsystematic, giving rise to a high entropy system. Moreover, each of these two systems have different light profiles. Spiral galaxies tend to be less concentrated and have exponential profiles; ellipticals have more of their light concentrated near their center and most bright ellipticals have de Vaucouleurs profiles. We find that the dimmer galaxies, which have presumably experienced fewer mergers, are mostly well described by exponential profiles. For galaxies in the middle of our luminosity range, the environment is strongly correlated with profile type. Galaxies in high density environments experience, on average, more mergers than those in low density environments. The brightest galaxies, and presumably the ones that have been created through many mergers, are predominantly de Vaucouleurs.

The dependences of profile type and apparent axis ratio on environment show statistically interesting trends on scales up to 2 $h^{-1}$ Mpc. This is consistent with other studies of the correlations of galactic properties with environment \citep{bl04}. The properties of a galaxy are generally affected by the presence of other galaxies in the same group or cluster, but not by galaxies
in neighboring clusters.

\begin{figure}
\plotone{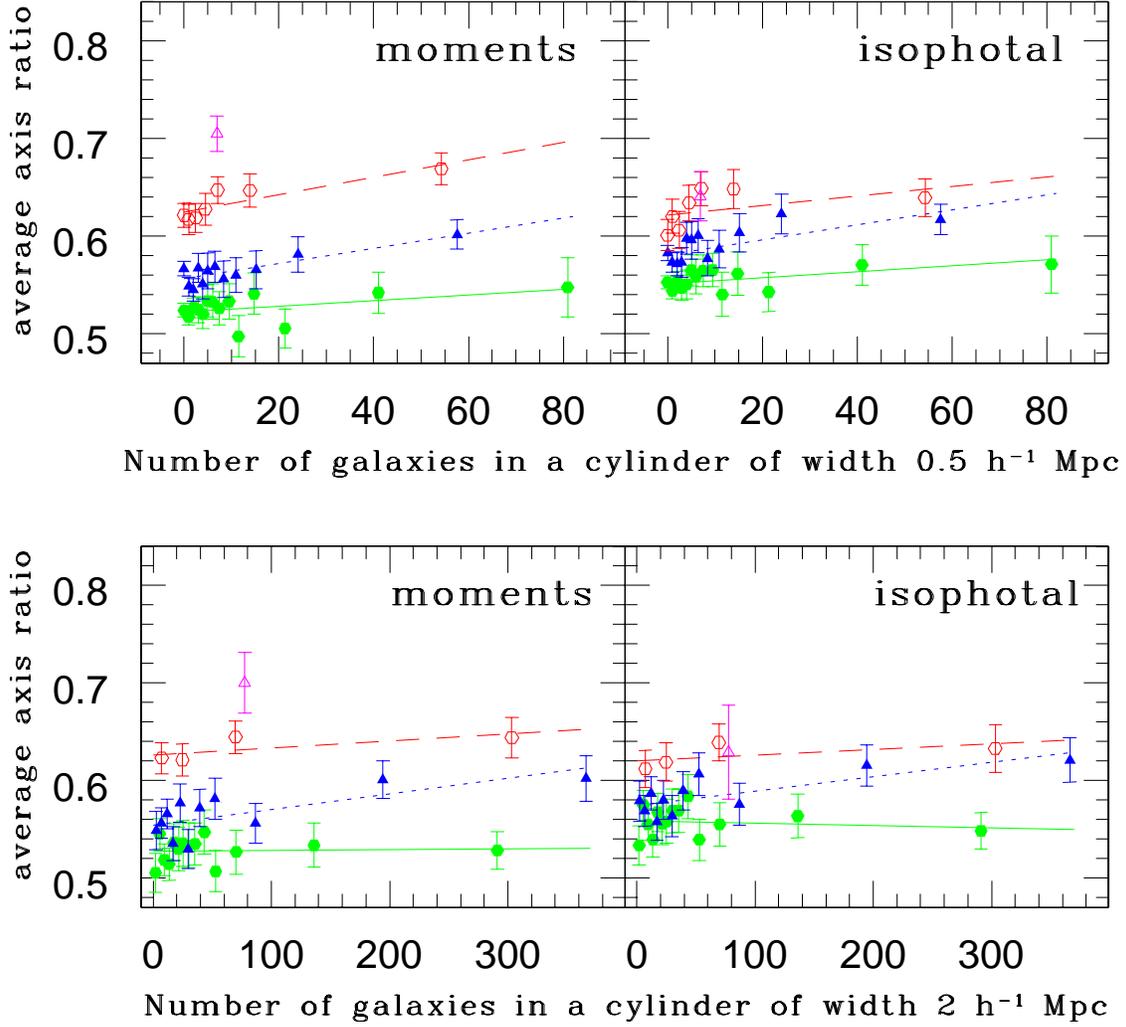}
\caption{Apparent axis ratio -- environment relation, using target galaxies in the absolute magnitude range $M_r\in [-18, -20]$ and cylinders of width 0.5 $h^{-1}$ Mpc (top panel) and 2 $h^{-1}$ Mpc (bottom panel) . Straight lines are $\chi^2$ fits to the binned data. Left panel plots are for the axis ratio as derived by the moments of the light distribution; right panel plots are for the axis ratio derived from the 25 mag arcsec$^{-2}$ isophote. The line types, colors, and point types are identical to those in figure \ref{fig:2}.
}
\label{fig:5}
\end{figure}

\begin{figure}\epsscale{0.9}
\plotone{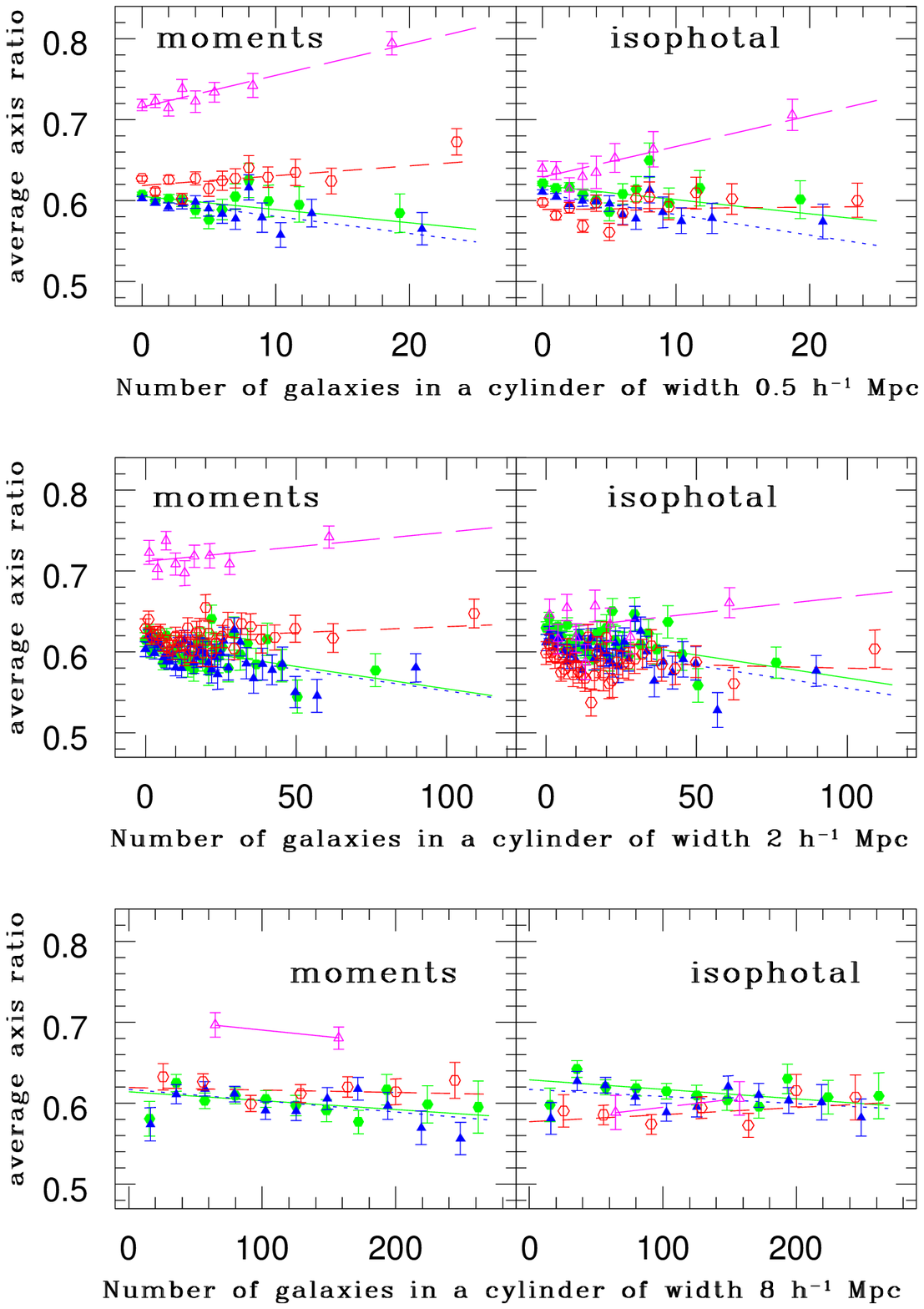}
\caption{Same as Figure~\ref{fig:5}, but with target galaxies in the absolute magnitude range $M_r\in [-20, -22]$ and cylinders of width 0.5 $h^{-1}$ Mpc (top panel), 2 $h^{-1}$ Mpc (middle panel) and  8 $h^{-1}$ Mpc (bottom panel).
}
\label{fig:6}
\end{figure}

\begin{figure}
\plotone{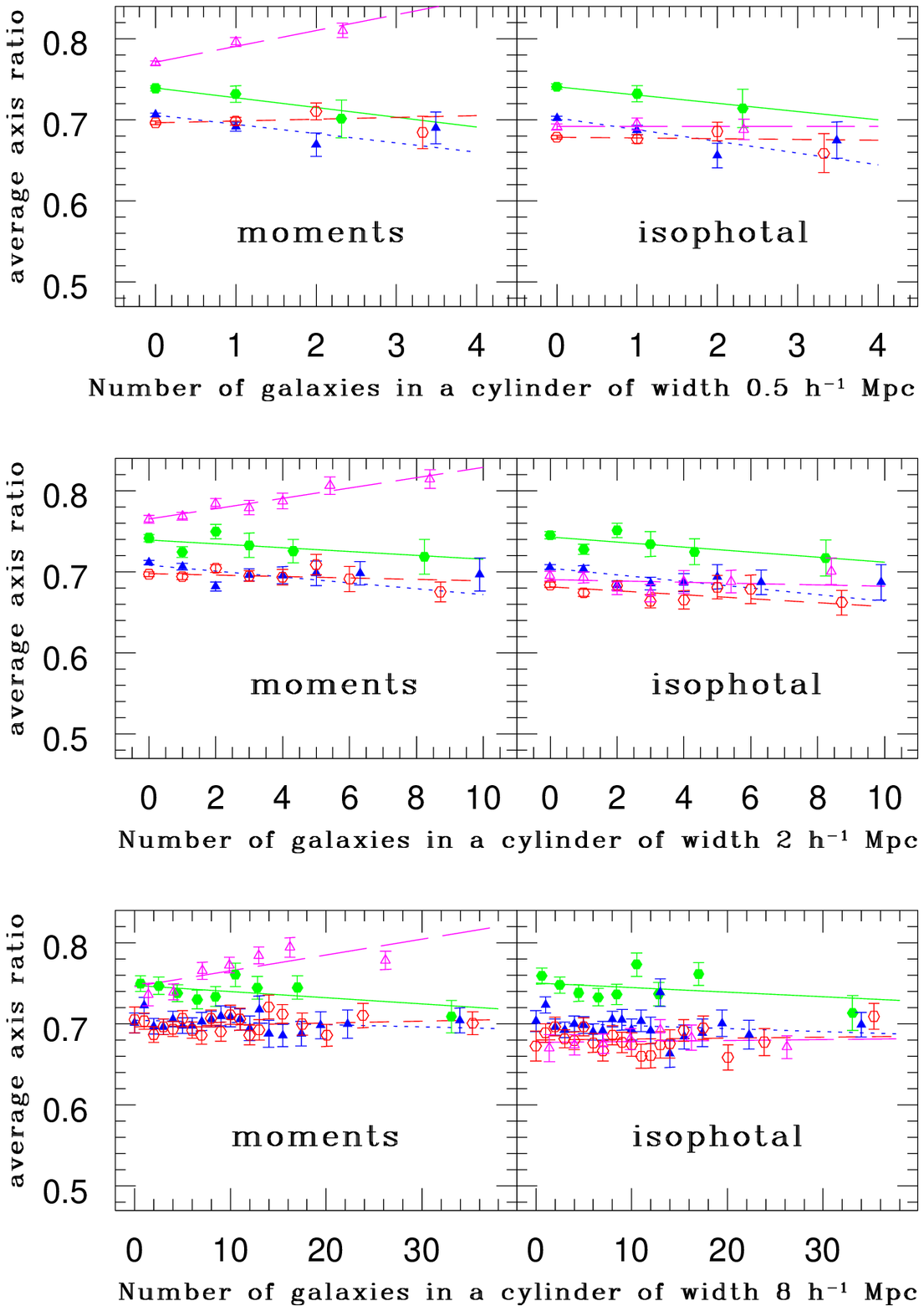}
\caption{Same as Figure~\ref{fig:5}, but with target galaxies in the absolute magnitude range magnitude range $M_r\in[-20, -22]$ and cylinders of width 0.5 $h^{-1}$ Mpc (top panel), 2 $h^{-1}$ Mpc (middle panel) and  8 $h^{-1}$ Mpc (bottom panel).
}
\label{fig:7}
\end{figure}

\citet{la92} use the APM Bright Galaxy Survey \citep{ma90} and study the axis ratio - environment relationship for elliptical galaxies. They arrive at the conclusion that ``local density is not an important factor in determining the flattening of ellipticals''. The lack of redshift information constrains \citet{la92} to estimate the local density in projection. This dilutes the signal by inevitably counting galaxies that have similar celestial coordinates as neighbors, while in reality they are separated by a large distance in space. Furthermore, they use a fixed angular scale on the sky within which they count the number of neighbors. For nearby galaxies, this scale corresponds to a larger physical distance than for distant galaxies. Our plots (Figures \ref{fig:5} through \ref{fig:7}) show that trends tend to get washed out as cylinder size increases. They are effectively mixing different cylinder sizes and in the process washing out any trend. Within this angular area, \citet{la92} label their galaxies as `high-density' if there are more than two neighboring galaxies, and `low-density' if there are two or less. This cut divides their sample into an approximately equal number of high and low density galaxies. They claim that there is no statistical difference in the observed distribution of their high and low density ellipticals. We performed a Komolgorov-Smirnov test on the distribution of axis ratios derived from the moments, for the \emph{de} galaxies with $M_r\in[-20,-22]$ in $0.5\, h^{-1}$ Mpc cylinders. The KS probability is minimized when we place the cut between low and high densities at eight nearest neighbors; this results in a probability of $P_{KS}=7\times 10^{-8}$. Instead of dividing our sample into two equal parts, this cut corresponds to a low/high density split of $\sim9/1$ . Thus \citet{la92} did not see the trend for two reasons: density in projection is a poor measure, and the difference in shape is most significant in very high density environments.

One open problem is whether or not field ellipticals are created and evolve in a fundamentally different way from those found in high density environments. \citet{da05} have shown that the massive early type galaxies in low density environments appear to have younger stellar populations and be slightly more metal rich than the ones in high density environments. \citet{rf01} have shown that elliptical galaxies with younger stellar populations tend to be flatter in their inner regions than ones with older stellar populations. One possible explanation for such behavior is that the central black hole in older ellipticals has had a longer time to randomize the orbits of stars near the center. In light of our result that the inner regions of \textit{de} galaxies tend to be more flattened in low density versus high density regions, these two previous studies may point to a more fundamental cause for our observed trends. However, using an age -- environment correlation to explain our results is no more fundamental than simply saying that axis ratio and environment are correlated. Instead, the missing link is in understanding what processes form centrally concentrated galaxies such as massive ellipticals. 

\citet{bo05} have used numerical simulations to show that it is possible to form elliptical systems from the merger of a large disk with multiple smaller disks. When a large disk galaxy is merged with one or two small disks (with the mass ratio of the large disk to the small disk being 1:7) it is found that a galaxy of an intermediate type is created. Such a galaxy may correspond to the \textit{ex/de} and \textit{de/ex} galaxies observed in the SDSS. A well known fact is that the merger of two spirals of roughly equal mass leads to the creation of an elliptical (e.g. \citet{ba92}; \citet{he04}; \citet{bo05}). However, it has not been shown that the most common way of creating an elliptical is through the merger of two spirals. Other numerical studies \citep{he04} claim to find evidence that there exist elliptical galaxies that appear not to have been formed by a major merger of spirals. Such studies show that different merger histories can lead to the creation of a galaxy with a de Vaucouleurs profile. Moreover, different merger histories lead to galaxies with different intrinsic axis ratios \citep{ba92}. If it was determined conclusively that certain types of mergers are favored in high versus low density regions, it would be possible to deduce the average axis ratio of galaxies in each environment. Conversely, if different merger histories produce characteristically different axis ratios, it may be possible to work backwards and deduce which merger histories are more common in a given environment. In any case, the axis ratio of a galaxy, in its outer regions, could also depend on harassment from neighboring galaxies. However,  without knowing which merger histories are favored in different environments, and assuming that the observed environment is similar to the pre-merger environment, it is not possible to determine the `initial' axis ratio of a galaxy and thus separate the the effect of harassment from that of the merger history.

One intriguing fact uncovered is that the \textit{ex} and \textit{ex/de}
galaxies with $M_r \sim -20$ are flatter, on average, in denser environments.
These galaxies are primarily flattened disk galaxies, with a past free
from major mergers. In a high density environment, a galaxy may be
harassed by numerous encounters in the past, or may have a tidal tail from
a recent close encounter. These processes may distort the apparent
shape of galaxies in dense regions. However, we visually inspected
all the \textit{ex} and \textit{ex/de} galaxies that have 25 or
more neighboring galaxies in a cylinder of radius $0.5 h^{-1}$ Mpc.
There were 35 such galaxies, of which only one appeared to have
a tidal tail and only one was a possible recent merger remnant.
We also inspected 35 \textit{ex} and \textit{ex/de} galaxies with
at most one neighbor; none of them had obvious tidal tails or
asymmetric distortions. Almost all galaxies in our visual samples
were separated from their neighbors by a distance significantly
larger than the galaxy's visual radius. If the greater average
flattening at higher density is due to tidal effects, then the
tidal distortions must be subtle, rather than taking the form of
strong tidal tails.

The SDSS has recently been used to study the correlation of many galactic properties with each other. Exhaustive studies, such as those produced by \citet{bl03}, \citet{hg03} and \citet{hg04}, arrive at the conclusion that luminosity and color are the fundamental galactic properties from which all correlations of other properties with environment are derived. Such claims are tested in and are in reasonable agreement with numerical simulations as described, for example, in \citet{bm04}. However, we find a trend, namely the apparent axis ratio -- local galaxy density correlation, for galaxies of the same \textit{r} band luminosity and surface brightness profile type, that is not due to a correlation of luminosity or color with environment.

We thank Andreas Berlind and the anonymous referee for useful discussion 
and feedback.

Funding for the creation and distribution of the SDSS Archive has been
provided by the Alfred P. Sloan Foundation, the Participating Institutions,
the National Aeronautics and Space Administration, the National Science
Foundation, the U.S. Department of Energy, the Japanese Monbukagakusho,
and the Max Planck Society. The SDSS website is \url{http://www.sdss.org/}.
The SDSS is managed by the Astrophysical Research Consortium (ARC) for
the Participating Institutions. The Participating Institutions are
The University of Chicago, Fermilab, the Institute for Advanced Study,
the Japan Participation Group, The Johns Hopkins University, Los Alamos
National Laboratory, the Max-Planck-Institute for Astronomy (MPIA), the
Max-Planck-Institute for Astrophysics (MPA), New Mexico State University,
University of Pittsburgh, Princeton University, the United States Naval
Observatory, and the University of Washington.

\begin{deluxetable}{lrccccrrr}
\tablewidth{0pt}
\tablecaption{Galaxy Profile Types: Binned by Luminosity\label{tab:profile}}
\tablehead{
\colhead{Luminosity} & \colhead{Total} & \multicolumn{4}{c}{Fraction of each profile type} & \multicolumn{3}{c}{Cylinder width}\\ 
\colhead{Bin Width}& \colhead{Galaxies} & \colhead{\textit{ex}} &
\colhead{\textit{ex/de}} & \colhead{\textit{de/ex}} & \colhead{\textit{de}} & \colhead{$0.5\, h^{-1} {\rm\,Mpc}$} & \colhead{$2\, h^{-1} {\rm\,Mpc}$} & \colhead{$8\, h^{-1} {\rm\,Mpc}$} }
\startdata
$-17 \to -18$ & 1610 & 0.56 & 0.30 & 0.11 & 0.03 & 1140 & 253 & \ldots \\
$-18 \to -19$ & 5915 & 0.50 & 0.31 & 0.14 & 0.05 & 5267 & 2410 & \ldots \\
$-19 \to -20$ & 15580 & 0.41 & 0.30 & 0.17 & 0.12 & 14698 & 9192 & \ldots \\
$-20 \to -21$ & 46755 & 0.27 & 0.25 & 0.20 & 0.28 & 45549 & 36208 & 8856 \\
$-21 \to -22$ & 81037 & 0.12 & 0.20 & 0.22 & 0.46 & 79988 & 70560 & 26975 \\
$-22 \to -23$ & 57900 & 0.04 & 0.14 & 0.21 & 0.61 & 57482 & 54219 & 30614 \\
\enddata
\end{deluxetable}

\begin{deluxetable}{lrccccrrr}
\tablewidth{0pt}
\tablecaption{Galaxy Axis Ratio: Binned by Luminosity\label{tab:shape}}
\tablehead{
\colhead{Luminosity} & \colhead{Total} & \multicolumn{4}{c}{Fraction of each profile type} & \multicolumn{3}{c}{Cylinder width}\\ 
\colhead{Bin Width}& \colhead{Galaxies} & \colhead{\textit{ex}} &
\colhead{\textit{ex/de}} & \colhead{\textit{de/ex}} & \colhead{\textit{de}} & \colhead{$0.5\, h^{-1} {\rm\,Mpc}$} & \colhead{$2\, h^{-1} {\rm\,Mpc}$} & \colhead{$8\, h^{-1} {\rm\,Mpc}$} }
\startdata
$-18 \to -20$ & 8669 & 0.53 & 0.35 & 0.11 & 0.01 & 7691 & 3459 & \ldots \\
$-20 \to -22 $ & 39964 & 0.37 & 0.39 & 0.20 & 0.04 & 38901 & 30536 & 7180 \\
$-22 \to -\infty $ & 20211 & 0.12 & 0.37 & 0.37 & 0.14 & 20055 & 18731 & 10068 \\
\enddata
\end{deluxetable}

\begin{deluxetable}{cccccccc}
\tablewidth{0pt}
\tablecaption{Galaxy Axis Ratio: Galaxies with $M_r \in [-18,-20]$\label{tab:shape18}}
\tablehead{
\colhead{Profile}&\colhead{Width} & \multicolumn{3}{c}{Isophotal}& \multicolumn{3}{c}{Moments}\\
\colhead{Type} & \colhead{($h^{-1}$ Mpc)} & \colhead{$\chi^2$/DOF} &\colhead{trend} & \colhead{$\sigma$} & \colhead{$\chi^2$/DOF} &\colhead{trend} & \colhead{$\sigma$} 
}
\startdata
\textit{ex}  & 0.5 & 0.28 & + & 1.1 & 0.38 & + & 0.96\\
\textit{ex/de}  & 0.5 & 0.50 & + & 2.8 & 0.43 & + & 2.9\\
\textit{de/ex} & 0.5 & 0.82 & + & 1.2 & 0.35  & + & 2.7\\
\hline
\textit{ex}  & 2 & 0.49 & - & 0.36 & 0.41 & + & 0.11\\
\textit{ex/de} & 2 & 0.40 & + & 2.4  & 0.67 & + & 2.6\\
\textit{de/ex} & 2 & 0.22 & + & 0.58 & 0.22 & + & 0.86\\
\enddata
\end{deluxetable}

\begin{deluxetable}{cccccccc}
\tablewidth{0pt}
\tablecaption{Galaxy Axis Ratio: Galaxies with $M_r \in [-20,-22]$\label{tab:shape20}}
\tablehead{
\colhead{Profile}&\colhead{Width} & \multicolumn{3}{c}{Isophotal}& \multicolumn{3}{c}{Moments}\\
\colhead{Type} & \colhead{($h^{-1}$ Mpc)} & \colhead{$\chi^2$/DOF} &\colhead{trend} & \colhead{$\sigma$} & \colhead{$\chi^2$/DOF} &\colhead{trend} & \colhead{$\sigma$} 
 }
\startdata
\textit{ex} & 0.5 & 1.2 & - & 2.5& 0.83 & - & 2.2\\
\textit{ex/de} & 0.5 & 0.66 & - & 4.2 & 0.74 & - & 3.7\\
\textit{de/ex} & 0.5 & 1.9 & + & 0.23 & 2.0 & + & 2.4 \\
\textit{de} & 0.5 & 0.59 & + & 3.6 & 0.34 & + & 4.9\\
\hline
\textit{ex} & 2 & 1.4 & - & 3.6 & 1.4 & - & 3.4\\
\textit{ex/de} & 2 & 1.3 & - & 4.0 & 1.2 & - & 4.0\\
\textit{de/ex} & 2 & 1.1 & - & 0.52 & 0.93 & + & 1.2\\
\textit{de} & 2 & 1.3 & + & 1.1 & 0.93 & + & 1.4\\
\hline
\textit{ex} & 8 & 0.75 & - & 1.7 & 0.76 & - & 1.6\\
\textit{ex/de} & 8 & 1.0 & - & 1.3 & 1.2 & - & 2.1\\
\textit{de/ex} & 8 & 0.61 & + & 0.88 & 0.77 & - & 0.37\\
\enddata
\end{deluxetable}


\begin{deluxetable}{cccccccc}
\tablewidth{0pt}
\tablecaption{Galaxy Axis Ratio: Galaxies with $M \in [-22,-\infty]$\label{tab:shape22}}
\tablehead{
\colhead{Profile}&\colhead{Width} & \multicolumn{3}{c}{Isophotal}& \multicolumn{3}{c}{Moments}\\
\colhead{Type} & \colhead{($h^{-1}$ Mpc)} & \colhead{$\chi^2$/DOF} &\colhead{trend} & \colhead{$\sigma$} & \colhead{$\chi^2$/DOF} &\colhead{trend} & \colhead{$\sigma$}
 }
\startdata
\textit{ex} & 0.5 & 0.012 & - & 1.3 & 0.13 & - & 1.6\\
\textit{ex/de} & 0.5 & 0.60 & - & 3.6  & 0.66 & - & 3.1\\
\textit{de/ex} & 0.5 & 0.29 & + & 0.23 & 0.45 & + & 0.67\\
\textit{de} & 0.5 & 0.0010 & + & $6.0\times 10^{-4}$ & 0.55 & + & 5.5 \\
\hline
\textit{ex} & 2 & 1.0 & - & 1.4 & 0.99 & - & 0.96\\
\textit{ex/de} & 2 & 0.75 & - & 3.5 & 1.3 & - & 3.3\\
\textit{de/ex} & 2 & 0.86 & - & 2.1 & 0.65 & - & 0.83\\
\textit{de} & 2 & 0.35  & - & 0.86 & 0.45 & + & 5.0\\
\hline
\textit{ex} & 8 & 1.0  & - & 1.1 & 0.57 & - & 1.1 \\
\textit{ex/de} & 8 & 0.91 & - & 1.5 & 0.56 & - & 1.2\\
\textit{de/ex} & 8 & 1.2 & + & 0.40 & 0.85 & + & 0.25\\
\textit{de} & 8 & 0.38 & + & 0.35 & 1.8 & + & 2.4\\
\enddata
\end{deluxetable}


\newpage

\begin{thebibliography}{}

\bibitem[Abadi et al.(2003)]{ad03}
Abadi, M. G., Navarro, J. F., Steinmetz, M., \& Eke, V. R. 2003, \apj, 591, 499

\bibitem[Abazajian et al.(2003)]{ab03}
Abazajian, K., et al. 2003, AJ, 126, 2081

\bibitem[Abazajian et al.(2004)]{ab04}
Abazajian, K. et al. 2004, AJ, 128, 502

\bibitem[Abazajian et al.(2005)]{ab05}
Abazajian, K. et al. 2005, AJ, submitted (astro-ph/0410239)

\bibitem[Adelman-McCarthy et al.(2005)]{am05}
Adelman-McCarthy, J. K. et al. 2005 astro-ph/0507711

\bibitem[Barnes(1992)]{ba92}
Barnes, J. E. 1992, ApJ, 393, 484

\bibitem[Bell et al.(2005)]{be05}
Bell, E. et al., 2005, astro-ph/0506425

\bibitem[Berlind et al.(2004)]{bm04}
Berlind, A. A., et al. 2004, astro-ph/0406633

\bibitem[Bernstein \& Jarvis(2002)]{bj02}
Bernstein, G. M., \& Jarvis, M. 2002, AJ, 123, 583

\bibitem[Bertschinger(1998)]{be98}
Bertschinger, E. 1998, ARAA, 36, 599

\bibitem[Binney \& de Vaucouleurs(1981)]{bv81}
Binney, J., de Vaucouleurs, G. 1981, \mnras, 194, 679 

\bibitem[Blanton et al.(2004)]{bl04}
Blanton, M. R., Eisenstein, D., Hogg, D. W., Zehavi, I. 2004, astro-ph/0411037

\bibitem[Blanton et al.(2003a)]{bl03}
Blanton, M. R., et al. 2003a, ApJ, 594, 186

\bibitem[Blanton et al.(2003b)]{blli03}
Blanton, M. R., Lin, H., Lupton, R., Maley, F. M., Young, N., Zehavi, I., \& Loveday, J. 2003b, AJ, 125. 2276

\bibitem[Bournaud et al.(2005)]{bo05}
Bournaud, F., Jog, C. J., \& Combes, F. 2005, astro-ph/0503189

\bibitem[de Vaucouleurs(1948)]{de48}
de Vaucouleurs, G. 1948, Ann. d'Astrophys. 11, 247

\bibitem[de Vaucouleurs(1959)]{de59}
de Vaucouleurs, G. 1959, Handb. Phys. 53, 275

\bibitem[Dressler(1980)]{dr80}
Dressler, A., 1980, \apj, 236, 351

\bibitem[Elmegreen et al.(2004)]{el04}
Elmegreen, D. M., Elmegreen, B. G., Hirst, A. C. 2004, \apj, 604, L21

\bibitem[Freeman(1970)]{fr70}
Freeman, K. C. 1970, ApJ, 160, 811

\bibitem[Fukugita et al.(1996)]{fu96}
Fukugita, M., Ichikawa, T., Gunn, J. E., Doi, M., Shimasaku, K.,
\& Schneider, D. P. 1996, AJ, 111, 1748

\bibitem[Goto et al.(2003)]{gc03}
Goto, T., 2003, MNRAS, 346, 601

\bibitem[Gunn et al.(1998)]{gu98}
Gunn, J. E., et al. 1998, AJ, 116, 3040

\bibitem[Hernandez et al.(2004)]{he04}
Hernandez, X., \& Lee, H. W. 2004, \mnras, 347, 1304 

\bibitem[Hirata \& Seljak(2003)]{hs03}
Hirata, C., \& Seljak, U. 2003, MNRAS, 343, 459

\bibitem[Hogg et al.(2003)]{hg03}
Hogg, D. W., et al. 2003, \apj, 585, L5

\bibitem[Hogg et al.(2004)]{hg04}
Hogg, D. W., et al. 2004, \apj, 601, L29

\bibitem[Hubble(1926)]{hu26}
Hubble, E. 1926, ApJ, 64, 321

\bibitem[Hubble \& Humason(1931)]{hh31}
Hubble, E., \& Humason, M. 1931, \apj, 74, 43

\bibitem[Lambas et al.(1992)]{la92}
Lambas, D. G., Maddox, S. J., \& Loveday, J. 1992, \mnras, 258, 404

\bibitem[Maddox et al.(1990)]{ma90}
Maddox, S. J., Sutherland, W. J., Efstathiou, G., Loveday, J. 1990, \mnras, 243, 692 

\bibitem[Naim et al.(1995)]{na95}
Naim, A., et al. 1995, MNRAS, 274, 1107

\bibitem[Nuijten et al.(2005)]{nu05}
Nuijten, M. J. H. M., Simard, L., Gwyn, S., R$\ddot{\mathrm{o}}$ttgering, H., J., A. 2005, \apj, 626, L77

\bibitem[Postman et al.(2005)]{pt05}
Postman, M., et al. 2005, \aj, 623, 721

\bibitem[Roberts \& Haynes(1994)]{rh94}
Roberts, M. S., \& Haynes, M. P. 1994, ARA\&A, 32, 115

\bibitem[Robertson et al.(2005)]{rb05}
Robertson, B., Hernquist, L., Bullock, J.S., Cox, T.J., Di Matteo, T., Springel, V., Yoshida, N. 2005, astro-ph/0503369

\bibitem[Ryden, Forbes, \& Terlevich(2001)]{rf01}
Ryden, B. S., Forbes, D. A., \& Terlevich, A. I. 2001, MNRAS, 326, 1141

\bibitem[Sandage et al.(1970)]{sa70}
Sandage, A., Freeman, K. C., \& Stokes, N. R. 1970 \apj, 160, 831

\bibitem[S\'ersic(1968)]{se68}
S\'ersic, J. L. 1968, Atlas de Galaxias Australes (Cordoba: Obs. Astron.)

\bibitem[Sheldon et al.(2004)]{sh04}
Sheldon, E. S., et al. 2004, \aj, 127, 2544

\bibitem[Smith et al.(2002)]{sm02}
Smith, J. A., et al. 2002, AJ, 123, 2121

\bibitem[Stoughton, et al.(2002)]{st02}
Stoughton, C., et al. 2002, AJ, 123, 485

\bibitem[Tanaka et al.(2004)]{ta04}
Tanaka, M., Goto, T., Okamura, S., Shimasaku, K., \& Brinkmann, J. 2004, \aj, 128, 2677

\bibitem[Thomas et al.(2005)]{da05}
Thomas, D., Maraston, C., Bender, R., \& De Oliveira, C. M. 2005, \apj, 621, 673  

\bibitem[Vincent \& Ryden(2005)]{vr05}
Vincent, R. A., \& Ryden, B. S. 2005, ApJ, 623, 137

\bibitem[York et al.(2000)]{yo00}
York, D. G. et al. 2000, AJ, 120, 1579

\end{thebibliography}
\end{document}